\documentclass[runningheads]{llncs}
\usepackage[T1]{fontenc}
\usepackage{csquotes}
\usepackage{environ}
\usepackage{booktabs}
\usepackage{graphicx}
\usepackage[hidelinks]{hyperref}
\usepackage{listings}
\usepackage{color}

\usepackage[acronym]{glossaries}

\newglossaryentry{projname}{name=RankAid,description={}}
\newacronym{api}{API}{application programming interface}
\newacronym{ai}{AI}{artificial intelligence}
\newacronym{dl}{DL}{deep learning}
\newacronym{htbg}{hTBG}{hierarchical Time-Biased Gain}
\newacronym{idcg}{IDCG}{ideal DCG}
\newacronym{llm}{LLM}{large language model}
\newacronym{mlp}{MLP}{multilayer perceptron}
\newacronym{ncf}{NCF}{Neural Collaborative Filtering}
\newacronym{ndcg}{NDCG}{Normalized Discounted Cumulative Gain}
\newacronym{nlp}{NLP}{natural language processing}
\newacronym{nux}{NUX}{Negative User Experience}
\newacronym{rs}{RS}{recommender system}
\newacronym{tmdb}{TMDB}{The Movie Database}

\begin{document}

\title{First, do no harm: Breaking suicidogenic echo chambers in media recommendation}
\titlerunning{Breaking suicidogenic echo chambers in RecSys}

\author{%
    Alberto Díaz-Álvarez\orcidID{0000-0002-4150-9052} \and
    Raúl Lara-Cabrera\orcidID{0000-0002-7959-1936} \and
    Fernando Ortega-Requena\orcidID{0000-0003-4765-1479} \and
    Víctor Ramos-Osuna\orcidID{0009-0009-6139-967X}
}
\authorrunning{A. Díaz-Álvarez et al.}

\institute{%
E.T.S.I. Sistemas Informáticos (Universidad Politécnica de Madrid), C. de Alan Turing, s/n, 28031 Madrid, Spain
\email{\{alberto.diaz,raul.lara,fernando.ortega,victor.ramos.osuna\}@upm.es}\\
\url{https://knodis-research-group.github.io/}
}

\maketitle

\begin{abstract}
\Acrlongpl{rs} generally optimises user engagement, but this approach is dangerous in mental health contexts. When vulnerable users show signs of suicidal ideation, standard algorithms often trap them in echo chambers of harmful content, worsening their psychological state. In response, we introduce \gls{projname}, a re-ranking method that prioritises clinical safety alongside predictive relevance. It works as an add-on layer to existing models: it penalises risky items and boosts therapeutic content depending on the user's current level of vulnerability. We evaluated this approach using the MovieLens 1M dataset, where items were semantically annotated for clinical risk and therapeutic value using \acrlongpl{llm}. Our simulations show that our algorithm successfully blocks the recommendation of harmful content during crisis peaks, actively reshaping the feed to support emotional de-escalation. Furthermore, this safety intervention only causes a controlled, acceptable drop in standard accuracy metrics like \acrshort{ndcg}. By using asymmetric hyperparameters, \gls{projname} also gives system administrators the flexibility to tune the severity of the intervention based on specific clinical guidelines.
\end{abstract}

\section{Introduction}

\Glspl{rs} are an essential component in digital platforms, typically designed to maximise user engagement and retention. Their effectiveness is traditionally evaluated by ranking quality metrics. In this work, we adopt the \gls{ndcg}~\cite{jarvelin2002cumulated} as our primary accuracy metric. This metric is widely used as it rewards systems that place the most relevant items at the top while penalising those further down the list.

However, because these models prioritise click probability, they assume all high-engagement content is suitable. In mental health and suicide prevention, this optimisation is dangerous. Previous research shows that vulnerable users experiencing suicidal ideation often fall into \enquote{digital rumination}~\cite{rogers2017rumination}. In such cases, traditional \gls{rs} risk trapping individuals in echo chambers of nihilistic or self-harming content, potentially worsening their mental state \cite{arendt2019effects}.

While some studies use heuristics or language models to mitigate harmful content~\cite{garcia2023detecting,oak2025re}, there is currently no mathematical function that combines clinical vulnerability with psychological risk to explicitly penalise exposure.

We propose a \enquote{preventive ethical adjustment} through \gls{projname}. This algorithm implements a re-ranking scoring function that modifies predictions in crisis contexts. By weighing the user's latent vulnerability against the item's psychological risk and therapeutic value, \gls{projname} mathematically penalises harmful recommendations for high-risk profiles (regardless of engagement potential) and promotes therapeutic content to help de-escalate the crisis (or at least, the rumination).

\section{Related work}

Recent advancements in \glspl{llm} have expanded the intersection of \gls{ai} and mental health; yet, recommendation algorithms still fall short in crisis contexts.

\subsection{Suicide risk detection}

Current literature primarily addresses suicide prevention through \gls{nlp} techniques. Many studies use \gls{dl} models, such as \textit{transformer} architectures, to identify suicidal ideation and depressive rumination. Specifically, analysing non-clinical texts (such as social media posts or search histories) is essential to assess psychological states in digital environments, even without access to medical records~\cite{coppersmith2016exploratory,calvo2017natural}.

This line of research has evolved into triage systems that estimate intervention urgency based on these signals. Notable in this area is the adaptation of metrics like \gls{htbg}, designed to quickly identify individuals at suicide risk when time is limited~\cite{shing2020prioritization}. However, this approach focuses exclusively on alerting clinical evaluators; it lacks mechanisms to assess the psychological risk of individual content items or to prevent users from being exposed to harmful material.

On the other hand, conversational agents (i.e., chatbots designed for crisis intervention) have shown positive results in providing early emotional support~\cite{fitzpatrick2017delivering}. The drawback is that both these chatbots and detection systems operate independently of the user's regular content consumption on the platform. Consequently, they fail to address a fundamental problem: the \gls{rs} itself might suggest harmful content that worsens the crisis.

\subsection{Echo chambers}

The formation of echo chambers~\cite{nguyen2014exploring,bakshy2015exposure} has recently been mathematically modeled. For example, \cite{chee2024harm} shows that when user preferences evolve, static penalties on problematic content can fail or even increase long-term harm. To mitigate this, the \enquote{trustworthy \acrlongpl{rs}} paradigm~\cite{ge2024survey} requires systems to be fair, explainable, and safe. Within this safety dimension, researchers have proposed filters for negative experiences~\cite{garcia2023detecting} or the use of \glspl{llm} to re-rank feeds and hide self-harm material~\cite{oak2025re}. Our work also utilises the semantic capabilities of \glspl{llm}.

However, unlike models that delegate re-ranking directly to a \gls{llm} (which is computationally expensive and/or opaque), we use the \gls{llm} beforehand to build a clinical knowledge matrix. While prior studies use static penalties, our approach treats \gls{llm} assessments as variables in a mathematical function. This determines rewards or penalties based on the user's vulnerability, ensuring protection aligned with digital wellbeing principles.

\subsection{Ethical algorithm modification}

A common strategy to avoid echo chambers is to force the algorithm to (also) optimise content diversity. Some systems attempt to break these bubbles by combining diversity and accuracy metrics~\cite{wang2025uncertain}. However, in the mental health domain, purely mathematical diversity metrics are insufficient. Because they optimise list heterogeneity without explicitly modelling clinical risk, a system could generate a topically diverse list that still includes explicit, implicit, or ambivalent self-harming content~\cite{valentine2023recommender,scherr2022social}. This risk is exacerbated by position bias: high-risk items appearing at the top of the list receive disproportionate user attention~\cite{klimashevskaia2024survey}. Even if these diverse items are harmless, introducing variety without ensuring relevance or clinical suitability is often perceived as mere noise~\cite{ziarani2021serendipity}.

Given the risks of unfiltered optimisation, recommendation objectives must be adapted for clinical contexts. Although modifying formulas for clinical reasons is uncommon, the literature on fairness in \glspl{rs} considers the recommendation space a scarce resource that must be distributed fairly and under certain constraints~\cite{singh2018fairness}. Following this principle, several authors have adapted classic metrics by introducing penalties to ensure the visibility of minority groups~\cite{gao2022fair,balagopalan2025s}. Essentially, the algorithm's objective is redefined: sacrificing some predictive accuracy is accepted in exchange for meeting (in principle) ethical criteria.

The tech industry has also begun to acknowledge these harms (caused by their own algorithms) and has implemented mechanisms to limit exposure to negative user experiences (\glspl{nux})~\cite{garcia2023detecting}. The problem is that social networks typically apply these limits as reactive \textit{a posteriori} filters (heuristic \textit{down-ranking}), rather than structurally integrating them into the mathematical formulas that train and guide the model from the ground up.

\section{Methodology}

Instead of designing a new predictive model from scratch, our approach uses a two-stage pipeline. First, a standard model generates relevance predictions. Second, our algorithm intervenes to recalculate each item’s value based on clinical parameters before display. An advantage of this strategy is that it operates independently of the underlying recommendation model, simply adding an extra layer to re-rank the proposed items.

\subsection{System architecture}

The workflow implemented in our experiment consists of the following steps:

\begin{enumerate}
    \item \textbf{Candidate item generation}. We use a standard algorithm based on \gls{ncf}~\cite{he2017neural}. The task is formulated as implicit feedback to predict the probability of user interaction. As explained in~\autoref{s:evaluation-metrics}, ratings are binarized so that a \gls{mlp} model assigns an affinity score, $rel_{i,u} \in [0,1]$, to each item. The model processes user and item IDs through $128$-dimensional embedding layers, which are concatenated and passed through Dense layers ($64$ and $32$ units) with Dropout to prevent overfitting. Finally, a sigmoid output layer generates the interaction prediction.
    \item \textbf{Re-ranking} (\gls{projname} algorithm). After the \gls{ncf} model provides the $rel_{i,u}$ probabilities, our algorithm evaluates each item and applies penalties or rewards based on the user's mental state directly to these scores.
    \item \textbf{Final top-$N$ generation}. Items are re-sorted in descending order according to the adjusted scores and presented to the user.
\end{enumerate}

\subsection{Mathematical formulation for \gls{projname} algorithm}
\label{sec:math-form}

Our proposal adjusts the original recommendation by reducing the weight of harmful content and increasing the weight of therapeutic content based on the user's current sensitivity. Formally, this is defined as a point-wise scoring function (\autoref{eq:rankaid}):

\begin{equation}
RankAid(i, u) = \max(0, rel_{i,u} - \alpha(v_u \cdot \rho_i) + \beta(v_u \cdot \tau_i))
\label{eq:rankaid}
\end{equation}

Where:

\begin{itemize}
\item $v_u \in [0, 1]$ represents \textbf{user vulnerability}, ranging from stability ($0$) to an extreme crisis state ($1$). Following \textit{Fluid Vulnerability Theory}~\cite{rudd2006fluid}, we treat risk as a dynamic state, assuming it is inferred via external methods (e.g., \gls{nlp} techniques).
\item $\rho_i \in [0, 1]$ represents the \textbf{risk factor}, indicating an item's toxicity or danger. We assume this is pre-labelled by automatic classifiers or specialists.
\item $\tau_i \in [0, 1]$ represents the \textbf{therapeutic rescue factor}, which measures an item's potential for emotional de-escalation (e.g., content that helps shift focus away from trauma).
\item $\alpha$ and $\beta$ are \textbf{intervention hyperparameters}. These weights control the severity of the risk penalty ($\alpha$) and the therapeutic reward ($\beta$). Using independent parameters instead of a convex combination (like $\lambda$ and $1 - \lambda$) allows for asymmetrical calibration. This gives clinicians the flexibility to adjust protection levels—such as applying strict censorship without limiting therapeutic content—rather than treating mitigation and promotion as a zero-sum game.
\end{itemize}

The function is designed so that if a user shows no clinical risk ($v_u \approx 0$), the penalty and reward terms disappear. This keeps the original relevance value $rel_{i,u}$ and ensures the recommender performs normally for the general population. However, as vulnerability increases ($v_u \rightarrow 1$), the term $-\alpha(v_u \cdot \rho_i)$ significantly reduces the score of harmful items, even if their $rel_{i,u}$ was high. Unlike traditional interventions (such as \textit{fairness}) that only act as penalties, our approach includes the term $+\beta(v_u \cdot \tau_i)$ that boosts items with therapeutic value.

\subsection{Evaluation metrics}
\label{s:evaluation-metrics}

To evaluate our algorithm, we used a standard 80/20 train-test split. We converted the explicit ratings into an implicit task by binarizing the labels: ratings above $4$ were treated as positive interactions ($1$). Unobserved items were then added using negative sampling at a 1:4 ratio to train the baseline \gls{ncf} model.

Accuracy is measured against the explicit ground truth of the $20\%$ test set using \gls{ndcg} (see \autoref{eq:ndcg}). This ensures that any measured degradation reflects the actual loss in ranking quality relative to the user's real preferences.

\begin{align}
    NDCG_p &= \frac{DCG_p}{IDCG_p} \\
    DCG_p  &= \sum_{i=1}^{p} \frac{rel_i}{\log_2(i+1)}
    \label{eq:ndcg}
\end{align}

\Gls{idcg} is calculated by sorting test items by their original explicit ratings (1 to 5) rather than the implicit labels used for training. This prevents ties and provides a more precise ideal baseline. If items have identical ratings (e.g., several 5s), their relative order does not affect the cumulative sum. This normalisation allows for consistent performance comparison across all users.

\section{Experiments and results}

To empirically validate this methodology, we designed a simulation using the standard MovieLens 1M dataset. Since there are no public datasets that combine movie viewing history with real clinical diagnoses (for privacy reasons), we adapted the environment using movie metadata as proxy variables. To simulate the inherent variability of a real catalogue and assign rigorous values to the content, we designed a semantic enrichment pipeline for the movie metadata. First, we cross-referenced the dataset identifiers with \gls{tmdb} \acrshort{api} to extract the full synopsis of each item.

Subsequently, we implemented an automated clinical inference using a \gls{llm} (specifically, the open-source 35B-parameter Qwen 3.5 model) running in a local environment. Through prompt engineering, we instructed the model to assume the role of an expert psychologist in suicide risk assessment and jointly analyse the title, synopsis, and tags of each movie. For each item, the \gls{llm} generated an evaluation returning three structural variables:

\begin{itemize}
    \item \textbf{Risk factor} ($\rho_i \in [0, 1]$): A continuous value quantifying the toxicity level or the movie's potential to induce negative rumination.
    \item \textbf{Therapeutic rescue factor} ($\tau_i \in [0, 1]$): A continuous value reflecting the content's potential to facilitate cognitive emotional de-escalation.
    \item \textbf{Clinical label}: A categorical classification assigned to one of three classes: Harmful, Therapeutic, or Neutral.
\end{itemize}

To ensure reproducibility, \autoref{tab:prompt} details the structural instructions (prompt) provided to the language model for clinical inference.

\begin{table}[ht]
    \centering
    \caption{Prompt provided to the \acrshort{llm} for semantic annotation}
    \label{tab:prompt}
\begin{lstlisting}[
        frame=tb,
        basicstyle=\scriptsize\ttfamily,
        breaklines=true,
        linewidth=\textwidth,
        xleftmargin=0pt,
    ]
You are an expert clinical psychologist specializing in media impact on
vulnerable individuals, specifically those experiencing suicidal
ideation or acute psychological crises.

Analyze the following movie metadata and assess two metrics:
1. rho (risk factor, float 0.0 to 1.0): How harmful, triggering, or
   toxic the movie could be. High risk includes themes of hopelessness,
   self-harm, severe trauma, nihilism, or explicit graphic violence.
2. tau (therapeutic rescue factor, float 0.0 to 1.0): The potential to
   generate emotional de-escalation, comfort, or hope. High rescue
   includes lighthearted comedy, positive overcoming of adversity,
   relaxing pacing, or deep emotional validation.

Based on rho and tau, assign a clinical label which MUST strictly be one
of: "Harmful", "Therapeutic", or "Neutral".

Title: {title}
Genres: {genres}
Synopsis: {overview}

OUTPUT FORMAT:
You must return ONLY a raw, valid JSON object. Do not include markdown
formatting, code blocks (do not use ```json), introductions, or
explanations. Adhere strictly to this schema:

{{
  "rho": <float>,
  "tau": <float>,
  "clinical_label": "<string>"
}}
\end{lstlisting}
\end{table}

This approach allows us to obtain $\rho_i$ and $\tau_i$ values based on a semantic analysis of the items' content, reflecting the variability found in a real recommendation catalogue. \autoref{tab:dataset-stats} shows the general distribution of these clinical labels across the entire dataset.

\begin{table}[ht]
\centering
\caption{Distribution of clinical labels and metric averages in the enriched dataset}
\label{tab:dataset-stats}
\begin{tabular}{lcccc}
\toprule
Clinical label & Count & Percentage ($\%$) & Avg. risk ($\rho$) & Avg. Rescue ($\tau$) \\
\midrule
Harmful & 701 & 18.0600 & 0.7893 & 0.1825 \\
Neutral & 1302 & 33.5400 & 0.3895 & 0.3564 \\
Therapeutic & 1879 & 48.4000 & 0.2111 & 0.7107 \\
\bottomrule
\end{tabular}
\end{table}

Additionally, \autoref{tab:llm-examples} provides concrete examples of the \gls{llm}'s semantic annotation. This demonstrates how the model interprets specific content to assign risk and rescue factors.

\begin{table}[ht]
\centering
\caption{Examples of LLM-based semantic annotation for clinical risk and therapeutic value}
\label{tab:llm-examples}
\begin{tabular}{lccc}
\toprule
Movie title & Risk ($\rho$) & Rescue ($\tau$) & Clinical label \\
\midrule
Copycat (1995) & 0.7000 & 0.3000 & Harmful \\
Assassins (1995) & 0.6500 & 0.2500 & Harmful \\
Toy Story (1995) & 0.0500 & 0.8500 & Therapeutic \\
Jumanji (1995) & 0.1000 & 0.8000 & Therapeutic \\
Heat (1995) & 0.4500 & 0.3500 & Neutral \\
Sudden Death (1995) & 0.4000 & 0.3000 & Neutral \\
\bottomrule
\end{tabular}
\end{table}

To establish a baseline for this experiment, we initialised the hyperparameters $\alpha$ and $\beta$ at $0.2$. As detailed in \autoref{sec:math-form}, these soft weights act as a starting \enquote{nudge} for the intervention, so they are designed to be adjusted based on the required domain/clinical criteria.

\subsection{Crisis escalation simulation}

With this structure, the experiment consists of simulating a progressive escalation in the user's crisis (increasing the $v_u$ variable from $0$ to $1$). The goal is to evaluate how the function changes the exposure of items in the top ranking positions compared to a traditional recommender.

\subsubsection{Exposure control}

\begin{figure}[ht]
    \centering
    \includegraphics[width=0.85\linewidth]{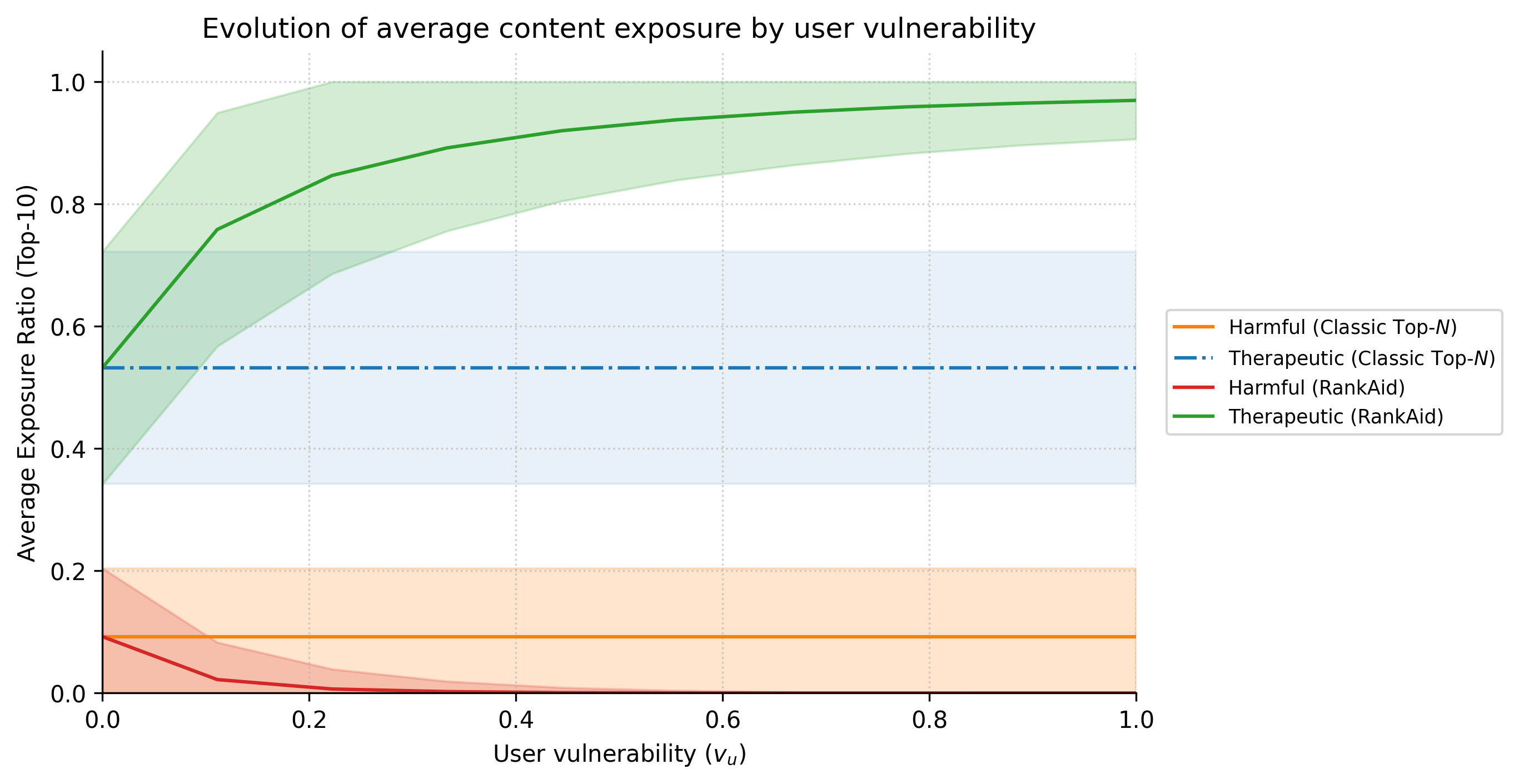}
    \caption{Average content exposure in the top-$10$ recommendations as user vulnerability ($v_u$) increases. While the classic model maintains constant exposure to harmful (orange) and therapeutic items (blue), \gls{projname} adapts dynamically. As the crisis escalates, it eliminates harmful content (red) and boosts therapeutic recommendations (green). Shaded areas show the standard deviation across all users.}
    \label{fig:exposure}
\end{figure}

Initially, we evaluated the clinical safety of the recommended items. As shown in \autoref{fig:exposure}, a classic \gls{rs} maintains a constant exposure to harmful content (around $10\%$) regardless of the user's state. In contrast, \gls{projname} significantly changes this dynamic.

As the crisis progresses, the recommender adapts through two protective behaviours. First, it mitigates harm: as vulnerability increases, harmful content decreases until it is completely removed from the top-$10$. Second, the system actively injects \enquote{rescue} content. Items with high therapeutic value ($\tau_i$), which represent about $55\%$ of recommendations during emotional stability ($v_u = 0$), increase to nearly $95\%$ during maximum crisis ($v_u = 1$). This dual mechanism demonstrates that the re-scoring function effectively avoids ruminative echo chambers and transforms the recommendation space into a de-escalation environment.

\subsubsection{(Controlled) degradation of accuracy}

Any deviation from an algorithm's original predictions inevitably reduces standard accuracy. By prioritizing clinical criteria, the system stops optimizing exclusively for historical engagement. To quantify this impact, we evaluated the resulting rankings using the \gls{ndcg} metric.

\begin{figure}[ht]
    \centering
    \includegraphics[width=0.6\linewidth]{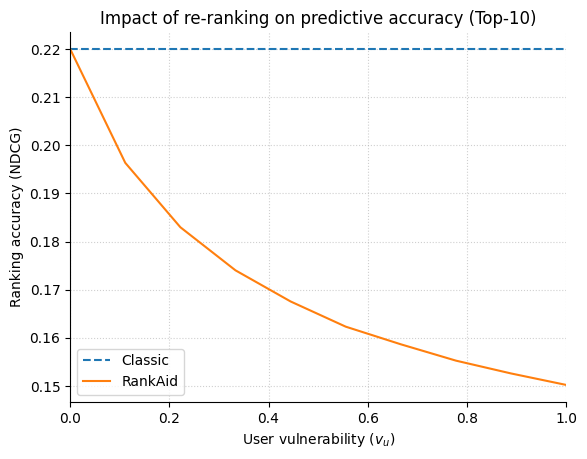}
    \caption{Controlled degradation of the \acrshort{ndcg} metric as the system prioritizes clinical safety. As the individual’s state requires more aggressive de-escalation, the algorithm moves away from recommendations based purely on interaction.}
    \label{fig:rankaid-in-dcg}
\end{figure}

As shown in \autoref{fig:rankaid-in-dcg}, \gls{projname} causes a slight downward trend in accuracy as user vulnerability increases. However, the results demonstrate that this decrease remains within acceptable margins, preserving a stable \gls{ndcg} index throughout the process.

From an ethical point of view, this reduction is not a model deficiency but a deliberate and acceptable \enquote{sacrifice}. The system accepts being slightly less accurate in predicting impulsive (and potentially ruminative) preferences in exchange for a safer digital environment that protects the user's psychological integrity. A detailed comparison of this trade-off at key vulnerability thresholds is presented in \autoref{tab:performance-snapshot}.

\begin{table}[ht]
\centering
\caption{Performance and safety comparison at key vulnerability thresholds}
\label{tab:performance-snapshot}
\begin{tabular}{lcccccc}
\toprule
Vulnerability ($v\_u$) & Model & NDCG & Harmful exp. ($\%$) & Rescue exp. ($\%$) \\
\midrule
0.00 & Classic Top-$N$ & 0.220 & 9.2 & - \\
0.00 & RankAid & 0.220 & 9.2 & 53.2 \\
0.56 & Classic Top-$N$ & 0.220 & 9.2 & - \\
0.56 & RankAid & 0.162 & 0.0 & 93.8 \\
1.00 & Classic Top-$N$ & 0.220 & 9.2 & - \\
1.00 & RankAid & 0.150 & 0.0 & 97.0 \\
\bottomrule
\end{tabular}
\end{table}

It is important to note that the absolute values of \acrshort{ndcg} for the base model are relatively low compared to the state of the art. The main objective of this experiment is not to maximise the predictive accuracy of the base model, but to demonstrate the relative dynamics of degradation and the safety mechanism of the \gls{projname} intervention. Because of this, the relative drop between the baseline and the intervened model is the relevant metric for our study, proving that our proposal is mathematically viable.

\subsection{$\alpha$ and $\beta$ calibration}

\begin{figure}[ht]
    \centering
    \includegraphics[width=0.6\linewidth]{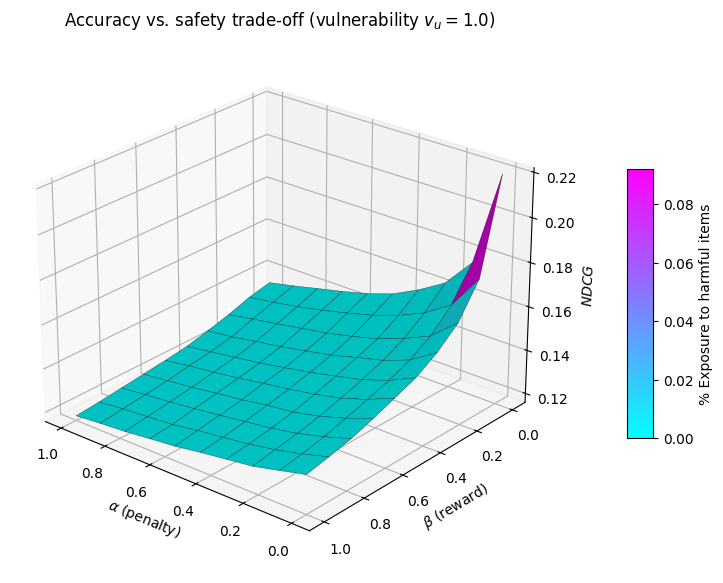}
    \caption{Pareto front showing the trade-off between accuracy (\acrshort{ndcg}) and safety (exposure to harm) for different combinations of $\alpha$ and $\beta$ during maximum crisis ($v_u = 1.0$).}
    \label{fig:alpha-beta-calibration}
\end{figure}

To understand how hyperparameters affect the recommender, we performed a grid search varying the risk penalty $\alpha$ and therapeutic reward $\beta$ between $0.0$ and $1.0$. Since the baseline probability $rel_{i,u}$ also falls within this range, these values are sufficient to either neutralize or maximally boost any item. The results are summarized in \autoref{fig:alpha-beta-calibration}.

The graph reveals a steep, non-linear surface that peaks when no risk or rescue factors are applied. However, even a tiny adjustment to $\alpha$ and $\beta$ suppresses risk while maintaining acceptable accuracy. These results demonstrate that risk suppression and rescue promotion operate separately, allowing for more flexible control.

\section{Conclusions}

To mitigate the formation of harmful echo chambers in scenarios of psychological vulnerability, we have proposed \gls{projname}, a re-ranking mechanism that balances predictive relevance with clinical safety. Our simulations demonstrate four key findings:

\begin{itemize}
    \item \textbf{Overcoming the limitations of previous approaches}. The simulations confirm that merely injecting diversity or using traditional fairness metrics is insufficient in clinical contexts. Unlike these methods, \gls{projname} does not just vary the content; it actively intervenes by neutralising risk and prioritising de-escalation.
    \item \textbf{Active de-escalation is viable (and efficient)}. The design of our scoring function is capable of autonomously intercepting echo chambers. By incorporating user vulnerability ($v_u$) as a dynamic variable, the system manages to reduce exposure to harmful content to zero at the peak of a crisis, replacing it with a set (nearly $100\%$) of alternatives with therapeutic value.
    \item \textbf{Viable accuracy-safety trade-off}. The slight predictive degradation (measured in \gls{ndcg}) is confirmed as an acceptable operational cost to ensure a psychologically safe digital environment.
    \item \textbf{Asymmetric hyperparameters}: The functional independence of $\alpha$ and $\beta$ provides the flexibility to calibrate the recommender's aggressiveness according to the clinical context, overcoming the limitations of zero-sum game approaches.
\end{itemize}

With our approach, we demonstrate that it is possible to modify recommenders so they stop being passive observers of user crises. Rather than applying general content filtering, our algorithm provides a personalised intervention. Under stable conditions, the system satisfies the user's natural preferences; however, when clinical risk indicators appear, \gls{projname} adapts the algorithm to act as a protective agent.

\begin{credits}
\subsubsection{\ackname}
This study is supported by the Dirección General de Investigación e Innovación Tecnológica de la Comunidad de Madrid (Orden 3177/2024) through the I+D Technological activities program (TEC-2024/COM-224).

\subsubsection{AI-assisted technologies disclosure}
During the preparation of this work, the authors used a \gls{llm} to assist with the translation and editing of the manuscript to improve language readability. After using this tool, the authors reviewed and edited the content as needed and take full responsibility for the content of the publication.

\subsubsection{Author Contributions}
The authors are listed in alphabetical order. All authors contributed equally to the research, development, and writing of this article.

\subsubsection{\discintname}
The authors have no relevant financial or non-financial interests to disclose.
\end{credits}

\bibliographystyle{splncs04}
\bibliography{references}

\end{document}